\documentstyle[revtex,eqsecnum]{aps}
\begin{document}
\draft
\columnsep -.375in
\twocolumn[
\begin{title}
Wentzel--Bardeen singularity and phase diagram for interacting\\
electrons coupled to acoustic phonons in one dimension\\
\end{title}
\author{Daniel Loss $^{1}$ and Thierry Martin $^{2*}$}
\begin{instit}

$^{1}$ Department of Physics, Simon Fraser University,
Burnaby, BC, Canada V5A1S6

$^{2}$ Theoretical Division, CNLS, Los Alamos National Laboratory,
Los Alamos, NM 87545
\end{instit}
\begin{abstract}
We consider strongly correlated electrons
coupled to low energy acoustic phonons in one dimension.
Using a Luttinger liquid description
we calculate the exponents of various response functions and discuss their
striking sensitivity to the Wentzel-Bardeen singularity
induced by the presence of phonons.
For the Hubbard model plus phonons the equivalent of a phase diagram is
established.
By increasing the filling factor towards half filling the
Wentzel--Bardeen singularity is approached. This in turn
triggers a simple and
efficient mechanism to suppress
antiferromagnetic fluctuations and to drive the system via a normal metallic
state towards a superconducting phase.
\end{abstract}
]
\narrowtext
Recently there has been a renewed interest in
one--dimensional (1D) interacting fermion systems,
both in the context of high temperature superconductivity
\cite{Anderson}, and
in the context of mesoscopic systems
\cite{Kane,Loss,LossM,Matveev}. 1D models are easier to
handle than their 2D and 3D
counterparts, and can yield useful information
for the latter systems. It has long been recognized
\cite{Mattis} that in 1D electron systems
the low-energy long wave--length
excitations around the Fermi surface
have bosonic properties.
The electron system is then represented as an harmonic fluid
\cite{Luther,Emery,Solyom},
called Luttinger liquid \cite{Haldane}, where the Fermi operators are
expressed in terms of scalar Bose fields.

In this letter, we  study
the properties of a 1D system of correlated electrons, interacting with
a short range repulsive potential, and coupled  to small-momentum transfer
bosons with acoustic dispersion. We concentrate first
on acoustic phonons, and later on discuss an application to electron
waveguides.
While the low dimensionality and the short range
forces rule out the possibility of
long range order at any finite temperature T, one can study
which type of  ordering fluctuations dominates the phase at vanishing
T\cite{Emery}
-- charge/spin density
waves (CDW/SDW) or singlet/triplet superconducting (SS/TS) fluctuations
-- and plot the equivalent of a phase diagram.

1D electron systems coupled to small-momentum phonons
were considered more than four decades ago by Wentzel\cite{Wentzel}
and  by Bardeen\cite{Bardeen} as candidates for the theory
of conventional superconductivity. It was pointed out
that for a critical phonon coupling
constant, the system becomes {\it unstable}, and the specific heat
diverges as one approaches this critical point \cite{Engelsberg}:
the retarded interaction
mediated by the phonons induces a collapse of the system.
Below, we refer to this singular point as the
Wentzel--Bardeen (WB) singularity.
Recently,
an exactly solvable model similar to that of Ref. \cite{Engelsberg} was
studied \cite{Voit0}, with the conclusion that
these small-momentum phonons can be ignored in typical metals since
their effect is of the order $c^2/v_F^2\ll 1$ ($v_F$ is the Fermi
velocity and $c$  the sound velocity).
Consequently, the interest \cite{Voit1,Zimanyi,Voit2}
shifted towards larger-momentum ($2k_F$) processes,
where an electron is scattered by a phonon across the Fermi surface.

However, we shall show that
the small-momentum phonons {\it do} play an
important role in strongly correlated systems,
as it is the velocity $u_\rho$ of particle--hole
excitations which determines
the ``adiabatic parameter'', $c^2/u_\rho^2$,
and not the Fermi velocity $v_F$.
In particular, we argue
in the following that:
1) the coupling to small-momentum transfer modes (such as acoustic phonons)
is a {\it non-perturbative} effect as one reaches the WB singularity.
2) For the Hubbard model, the WB singularity can be reached
for arbitrary electron--phonon
coupling constant by increasing the filling factor
towards half filling.
3) Near this WB singularity, CDW and SDW fluctuations are suppressed,
and the system is pushed into a metallic  and finally
superconducting phase by slightly increasing the filling factor.
Migdal's theorem
does therefore not apply in the presence of strong correlations.
Finally, it is worth pointing out that this high sensitivity
towards doping is reminiscent
of high temperature superconductivity materials.

We use the Luttinger liquid
description
generalized to electrons with spin \cite{Emery},
later on to be further
specified to a Hubbard model \cite{Schulz}.
Our starting point is the Hamiltonian,
$H=H_e+H_{ph}+H_{e-ph}$, where
\begin{equation}
H_{ph}={1\over 2}\int d x~[\zeta^{-1}\Pi_d^2+ \zeta c^2 (\partial_x
d)^2]
\label{phonon}\end{equation}
describes the free phonons, with $d$ the
displacement field, $\Pi_d$ its canonical
conjugate and $\zeta$ the mass density.
The electron--phonon coupling is
\cite{Fetter}:
\begin{equation}
H_{el-ph}=g\sqrt{\pi\over 2}\sum_s\int d
x~\psi_s^{\dagger}\psi_s \partial_x d~,
\label{coupling}\end{equation}
where $g$ is the coupling constant, and $\psi_s=
e^{ik_Fx}\psi_{s+}+e^{-ik_Fx}\psi_{s-}$
is the electron field operator. In terms of boson fields,
$\psi_{s\pm}({\bf x})\propto\exp\bigl[i\sqrt{\pi}\bigl(\pm \varphi_s({\bf
x})-\int^x\Pi_s(x^\prime,\tau)dx^\prime)\bigr)\bigr]$ (here ${\bf x}=(x,\tau$))
representing right and left moving electrons.
The electron density in Eq. (\ref{coupling}) reduces to $\sum_s\psi_s^\dagger
\psi_s=\sqrt{2/\pi}\partial_x\varphi_\rho$, with
$\varphi_{\rho,\sigma}=(\varphi_\uparrow\pm\varphi_\downarrow)/\sqrt{2}$
the charge and spin fields (similarly for the canonical
conjugate fields $\Pi_{\rho,\sigma}$). We have neglected the
fast oscillating terms in the density which give rise
to $2k_F$ phonon induced backscattering.
The electronic Hamiltonian is
$H_{el}=H_\rho+H_\sigma$, with
\begin{equation}
H_\nu={1\over 2}\int d x~\Biggl[u_\nu K_\nu\Pi_\nu^2
+{u_\nu \over K_\nu}(\partial_x\varphi_\nu)^2\Biggr]~,
\label{nu Hamiltonian}\end{equation}
with $\nu=\rho, \sigma$.
The phonons couple only to the charge degrees of freedom,
and this property is preserved in the propagators calculated
below. In Eq. (\ref{nu Hamiltonian}),
the electron interaction is included
in the parameters $K_\rho$, $K_\sigma$, and the charge
(spin) velocities $u_\rho$ ($u_\sigma$).
This model belongs to the same universality class
as the multicomponent Tomonaga--Luttinger model \cite{Penc}.

To calculate Green functions, we use
a Lagrangian formulation, where the partition function
is represented as a functional integral.
As we
will be calculating  Green functions which involve
fermion operators only, the phonon degrees of freedom are
integrated out right away. This leaves us with an effective action for
the charge degrees of freedom, which reads in  Fourier representation:
$S_\rho=(2\beta L)^{-1}\sum_{\bf k}D_\rho({\bf k})|\varphi_\rho({\bf
k})|^2$ (${\bf k}=(k,\omega)$), with the inverse propagator:
\begin{equation}
D_\rho({\bf k})={1\over K_\rho u_\rho}\biggl(\omega^2+u_\rho^2 k^2
-{b^2 k^4\over \omega^2+c^2k^2}\biggr)~,
\label{propagator}\end{equation}
where $b=g\sqrt{K_\rho u_\rho /\zeta}$. The first two terms
of Eq. (\ref{propagator})
represent the contribution of the free charge field, and the
retarded, attractive coupling associated with the phonons appears in
the third term.
At the WB point the charge density propagator,
$k^2/D_\rho({\bf k})$, becomes proportional
to $\omega^{-2}$ and signals an instability
towards long wave--length density fluctuations.

We first calculate the single--particle Green function:
\begin{equation}
G_s(x,\tau)=-<T\psi_s(x,\tau)\psi_s^\dagger(0,0)>~,
\label{Greens 0}\end{equation}
where $T$ is the imaginary time ordering operator.
This quantity has been calculated before \cite{Apostol}
(by diagrammatic methods)
for spinless fermions coupled to acoustic phonons, but
in the perturbative limit $c^2/v_F^2\ll 1$, where
the WB singularity is absent.
Using the decomposition of the Fermi operators into right and
left moving components, and after normal
ordering\cite{Loss,Martin}, the calculation of the Green function
is reduced to an evaluation of Gaussian integrals.
The result is ($a$ is the lattice constant):
\begin{eqnarray}
G_s(x,\tau)&=&{2\pi\over a}
\left|{a\over x+iu_\sigma\tau}\right|^{K_\sigma/4
+1/4K_\sigma}
\nonumber\\
&~&\times\prod_{\beta=\pm}\left|{a\over
x+iv_\beta\tau}\right|^{K_\rho
u_\rho F_\beta/4v_\beta+C_\beta/4K_\rho}\nonumber\\
&~&\times\sum_{\alpha=\pm}e^{i \alpha k_F x}
\biggl({s \alpha  x+iu_\sigma |\tau|\over
i|x+iu_\sigma \tau|}\biggr)^{sgn(\tau)/2}\nonumber\\
&~&\times\prod_{\gamma=\pm}
\biggl({ \alpha   x+iv_\gamma |\tau|\over
i|x+iv_\gamma \tau|}\biggr)^{sgn(\tau)F_\gamma /2}\nonumber\\
&~&\times n_\alpha (x,\tau)~,
\label{end result Greens}
\end{eqnarray}
with the normal ordering contribution:
\begin{equation}
n_\alpha(x,\tau)= {\rm sgn}(\tau)
\prod_{\nu=\rho,\sigma}e^{-\alpha\pi{\rm sgn}(\tau)
[\theta_\nu(x,\tau),\varphi_\nu(0,0)]/2}~,
\label{g n}\end{equation}
and thus $|n_\alpha|=1$. If $x/L\neq 0$ and $\tau$ arbitrary,
$n_\alpha (x,\tau)
={\rm sgn}(\tau)\exp(i\alpha\pi x{\rm sgn}(\tau)/L)$,
and if $\tau=0^{\pm}$ and $x$ arbitrary,
$n_\alpha(x,0^{\pm})=\pm i\alpha|x+i\alpha a|/(x+i\alpha a)$.
Near ${\bf x}=0$ one has to add
$a$ to $v|\tau|$  in Eq. (\ref{end result Greens}).
We introduced the velocities, $v_\pm^2=[u_\rho^2+c^2\mp\sqrt{
(u_\rho^2-c^2)^2+4b^2}]/2$, for the electronic collective excitations.
These velocities characterize the roots
of the propagator (\ref{propagator}), and
arise from the hybridization of the charge and phonon excitations.
Note that it is the charge velocity $u_\rho$
(not $v_F$) which appears here. The exponents are given by
$C_\pm=u_\rho(v_\pm^2 -c^2+b^2/u_\rho^2)/[(v_\pm^2-v_\mp^2)
v_\pm]$, and $F_\pm=(v_\pm^2-c^2)/(v_\pm^2-v_\mp^2)$.

At large distances, the Green function decays as a power law,
$G_s(x,0)\propto |x|^{-1-\delta}$,
with
$\delta=K_\sigma/4
+1/(4 K_\sigma)-1
+B/(4K_\rho)+A K_\rho/4$.
The electron--phonon parameters are defined by
$A= u_\rho(1+c^2/v_+v_-)/(v_++v_-) \ge 1$, and
$B= u_\rho(1+v_+v_-/u_\rho^2)/(v_++v_-)\le 1$.
These parameters play a crucial role in the discussion
of the ordering fluctuations below. In the limit $g\rightarrow 0$,
$A=B=1$. Let us now consider the case $g\neq 0$, where
$A>1$ and $B<1$\cite{Martin}:
As $u_\rho/K_\rho$
approaches the critical value
$u_\rho^*/K_\rho^*=g^2/(\zeta c^2)$ from above,
$v_+^2$ tends to zero, and $v_-^2$ to ${u_\rho^*}^2+c^2$.
As a result, the exponent $A$ diverges and  $B$ decreases
to the finite value $B^*=u_\rho^*/\sqrt{{u_\rho^*}^2+c^2}<1$.
For $u_\rho/K_\rho< u_\rho^*/K_\rho^*$ the velocity $v_+$ becomes
complex and the model becomes unphysical. Thus we must require
that $u_\rho/K_\rho\ge u_\rho^*/K_\rho^*$, or equivalently that
$b/(cu_\rho)\leq 1$, the equality sign defines the WB
singularity\cite{footnote1}.
This singular
behavior is a non-perturbative effect of the
electron-phonon coupling \cite{footnote2}
and originates from the instability of the propagator
(\ref{propagator}).
Because $\delta>0$, if $g\neq 0$ \cite{Martin}, the Green
function decays faster than $1/x$, the free
electron result. From (\ref{end result Greens}), we find the
momentum distribution function near $k_F$,
$N(k)\simeq N(k_F)-\kappa {\rm sgn}(k-k_F)|k-k_F|^{\delta}$,
with $\kappa$ some constant of order one, and
$N(k_F)=\Gamma(1/2+\delta/2)/[2\sqrt{\pi}\Gamma(1+\delta/2)]$.
This continuous but non-analytic behavior
at the Fermi surface is characteristic for Luttinger liquids
\cite{Mattis,Luther,Emery}.
The presence of phonons induces this Luttinger liquid
behavior even without  electron-electron interaction
($K_\rho=K_\sigma=1$).

We now turn to the discussion of the ordering fluctuations characterized
by two--particle response functions. The definitions for $N({\bf x})$,
$\chi({\bf x})$, $\Delta_s({\bf x})$ and $\Delta_t({\bf x})$
which describe CDW, SDW, singlet (SS) and triplet (TS)
superconducting fluctuations are given in \cite{Emery,Solyom}.
The calculation of these Green functions
will be given elsewhere
\cite{Martin}. Here, we give the final result in Matsubara representation:
\begin{mathletters}
\begin{eqnarray}
N(x,\tau)&\propto&\prod_{\beta=\pm}|x+iv_\beta\tau|^{-u_\rho K_\rho
F_\beta/v_\beta}|x+iu_\sigma \tau|^{-K_\sigma}\nonumber\\
\label{exponent N}\\
\chi(x,\tau)&\propto&\prod_{\beta=\pm}|x+iv_\beta\tau|^{-u_\rho K_\rho
F_\beta/v_\beta}|x+iu_\sigma \tau|^{-1/K_\sigma}\nonumber\\
\label{exponent chi}\\
\Delta_s(x,\tau)&\propto&\prod_{\beta=\pm}|x+iv_\beta\tau|^{-C_\beta/
K_\rho} |x+iu_\sigma \tau|^{-K_\sigma}
\label{exponent singlet}\\
\Delta_t(x,\tau)&\propto&\prod_{\beta=\pm}|x+iv_\beta\tau|^{-C_\beta/
K_\rho} |x+iu_\sigma \tau|^{-1/K_\sigma}~.
\label{exponent triplet}\end{eqnarray}
\end{mathletters}
The signature for a particular ordering fluctuation to be present is
given by the divergence of the Fourier transform of the corresponding
response function at low frequency and  small momentum
relative to $q=2k_F$ ($q=0$) for $N$ and $\chi$ ($\Delta_{s,t}$).
We thus obtain the following criteria\cite{Martin}:
\begin{mathletters}
\begin{eqnarray}K_\rho A+K_\sigma&\leq&
2~~~~~~~{\rm (CDW)~~~~}\label{criterion
C D W} \\
K_\rho A+1/K_\sigma&\leq&2~~~~~~~{\rm
(SDW)~~~~}\label{criterion S D W}\\
B/K_\rho+K_\sigma&\leq&2~~~~~~~{\rm (SS)}
\label{criterion singlet}\\
B/K_\rho+1/K_\sigma&\leq&2~.~~~~~~{\rm
(TS)}\label{criterion triplet}
\end{eqnarray}\end{mathletters}
Since $A>1$ and $B<1$ for $g\neq 0$, we see that the Cooper instability
is always present for non--interacting
electrons coupled to phonons. The exponents
can in principle be recovered in \cite{Penc},
which treats formally an arbitrary number
of coupled charge fields: here we have the advantage
of having explicit analytic expressions.

We now specialize to a Hubbard model for the correlated electrons.
In this case\cite{Schulz}, $K_\sigma=1$, and
the CDW and the SDW (the SS and the TS) response functions have
apparently the same exponents (except at half filling).
However, logarithmic corrections due to marginally
irrelevant operators in the spin channel favor SDW (TS)
over CDW (SS) fluctuations \cite{Logarithm}.
The remaining parameters $u_\rho$ and $K_\rho$ of the Luttinger
liquid Hamiltonian are obtained for arbitrary on--site repulsion
$U$ and filling factor $n$ by solving two integral equations
\cite{Schulz,Martin} describing the ground state properties \cite{Lieb}
and the spectrum of charge excitations \cite{Coll}.

Our results are plotted in Fig. \ref{fig1} a) and b):
for fixed $U/t$ (Fig. \ref{fig1} a), where $t$ is the bandwidth, we determine
which fluctuations dominate as a function of $n$ and an effective
electron--phonon parameter $b/(u_\rho c)$: for
convenience we  consider in these plots $n$ ($U$) and
$b/(u_\rho c)$ as independent parameters.
For small $b/(u_\rho c)$, SDW (i.e. antiferromagnetic) ordering fluctuations
dominate for
arbitrary filling factor. As this parameter is increased,
we reach (away from half filling) a region for which no
response function diverges: we refer to this ``phase''
as the metallic region. At low $U$ and
$b/(u_\rho c)\ll 1$ analytical results confirm the existence
of this intermediate phase \cite{Martin}.
By further increasing $b/(u_\rho c)$, the region where superconducting
fluctuations dominate is reached. On the other hand, near
half filling,
the correlation effects suppress the
superconducting phase.
(For larger values of $U$, the superconducting
region shrinks towards the region where $b/(u_\rho c)=1$,
and the SDW region grows as expected \cite{Martin}.)
Next, we choose quarter filling ($n=1/2$), and plot in Fig. \ref{fig1} b) the
phase diagram as a function of $U$ and the phonon coupling.
At low $U$, $K_\rho\approx 1$, and the system is superconducting
because of the Cooper instability. As $U$ is further
increased,  one crosses the metallic and the SDW (CDW) phase,
because the phonon--mediated attractive
interaction is overcome by the instantaneous repulsion between electrons.

In both Figs. \ref{fig1} a) and b), the upper line
$b/(u_\rho c)=1$ corresponds to the WB singularity. We now
discuss the relevance of this singularity for the Hubbard model.
Plotting $u_\rho/K_\rho$ for several values of $U$ as a function
of the filling factor (Fig. \ref{fig2}), we notice the well-known
fact\cite{Schulz}
that $u_\rho /K_\rho$
vanishes as one approaches half filling (and, of course, zero filling).
Thus, the WB singularity
at $u_\rho^*/K_\rho^*$ can be reached
{\it for arbitrary values of the coupling constant}
$g$ ($\neq 0$). Near half filling, as one approaches
$u_\rho^*/K_\rho^*$ from lower
filling factors, the divergence in $A$
triggers a dramatic suppression of the SDW (CDW) fluctuations,
and the system is driven into the metallic phase \cite{footnote}.
Moreover, since $B\rightarrow B^*$ as $u_\rho\rightarrow u_\rho^*$, and
$B^*\propto g^2$, the SS/TS condition $B\leq K_\rho$ can be met near
half filling for sufficiently small $g\neq 0$
(since $1/2\leq K_\rho\leq 1$).
Hence, the system can finally be driven into the superconducting phase
by approaching half filling.
In summary, the conjunction of the coupling to small momentum
acoustic modes with the reduction of the charge velocity near
half filling provides a very efficient mechanism to suppress SDW (CDW)
order, and to drive the system into a metallic and
finally superconducting phase. This mechanism
need not be limited to the
1D Hubbard model, but could occur in other situations
where strong correlations play an important role.


We conclude with an application to
mesoscopic quasi--1D wires.
The phonon field in Eqs.(\ref{phonon}) and (\ref{coupling})
now represents the charge excitations of
another transverse state.
This model was used in
\cite{Matveev} to describe the transport properties of a multimode wire
\cite{Larkin}.
The Fermi velocities associated with the two modes are comparable and
therefore the retarded attractive interaction
is much stronger than for phonons.
Beyond a critical
coupling, superconducting
fluctuations are induced
first in the branch with the lower Fermi velocity,
and then in both modes.
This could be probed e.g. by studying the periodicity
of the persistent current of a two channel mesoscopic ring in a
heterostructure \cite{Mailly}, as a function of electron density.
The flux causes a twist
in the boundary conditions, but the bulk properties
are unaffected.
The mutual inductance of the two modes is
negligible because of the small ratio $\alpha v_F/c_l$
($\alpha\sim 137$ and $c_l$ is the
velocity of light \cite{LossM}).
In the metallic regime,
the power spectrum
has a peak associated with the flux quantum
$\phi_0=hc/e$, whereas in the superconducting
case, this peak should give place to a peak
at the {\it superconducting} flux quantum
$\phi_0/2$

\acknowledgements
We thank A. J. Leggett, C.P. Enz, and S. Trugman for useful
discussions. The work of D.L. is supported by NSERC of Canada.

\figure{a) Phase diagram of the Hubbard model coupled to phonons,
as a function of filling factor and phonon coupling constant $b/cu_\rho$,
for the choices $U/t=0.3$ and $c=at$ (this corresponds e.g.to $(c/u_\rho)^2
\approx 1/2$ at quarter filling ($n=1/2$). $SDW$, $M$, and $SC$ label
the SDW, intermediate, and (triplet) superconducting regions.
b) Phase diagram of the Hubbard model coupled to phonons
at quarter filling ($n=1/2$), as a function of $U/t$ and $b/cu_\rho$.
\label{fig1}}
\figure{Plot of $u_\rho/K_\rho$ as a function of the filling factor $n$.
Note the abrupt change for small values of $U$ as one approaches half
filling.\label{fig2}}

\begin{references}
\bibitem{footnote1} This singularity was left unnoticed
in Ref. \cite{Voit0} (only forward scattering is includede there),
and the exponents given there seem to be incorrect.
\bibitem{footnote2}Note also that the limits $g\rightarrow 0$ and
$u_\rho\rightarrow u_\rho^*$ do not commute.
\bibitem{footnote} Note that the model becomes unphysical
when we go  below the WB singularity
$u_\rho^*/K_\rho^*$ in Fig. 2 by increasing (lowering)
the filling factor towards half (zero) filling.
\end{references}
\end{document}